\documentclass{elsart}
\pdfoutput=1
\newcommand{\avg}[1]{\left\langle{#1}\right\rangle}

\usepackage[ansinew]{inputenc} 
\usepackage[T1]{fontenc} 

\usepackage{graphics}

\usepackage{epsfig}

\usepackage{amssymb}
\usepackage{amsmath}

\begin{document}

\begin{frontmatter}



\title{How does informational heterogeneity affect the quality of forecasts?\thanksref{label1}}

 \thanks[label1]{We are indebted with J.D. Farmer for important comments and suggestions. ADM thanks the Santa Fe Institute for its kind hospitality.}


\author{S. Gualdi$^1$ and A. De Martino$^{1,2}$}

\address{$^1$Dipartimento di Fisica, Sapienza Universit\`a di Roma, P.le A. Moro 2, 00185 Roma, Italy\\
$^2$INFM(SMC)/CNR(ISC), Roma, Italy}


\begin{abstract}
We investigate a toy model of inductive interacting agents aiming to forecast a continuous, exogenous random variable $E$. Private information on $E$ is spread heterogeneously across agents. Herding turns out to be the preferred forecasting mechanism when heterogeneity is maximal. However in such conditions aggregating information efficiently is hard even in the presence of learning, as the herding ratio rises significantly above the efficient-market expectation of $1$ and remarkably close to the empirically observed values. We also study how different parameters (interaction range, learning rate, cost of information and score memory) may affect this scenario and improve efficiency in the hard phase.
\end{abstract}

\end{frontmatter}

\section{Introduction}

One of the most intriguing observations made over the years in financial data analysis concerns the tendency of financial forecasters to imitate each other. An account of this phenomenon is found in  \cite{bouchaud}, where earning forecasts from different markets over a 17 year period are analyzed, showing (among other results) that: (a) forecasts are typically optimistic, that is the difference between the forecast and the actual earning is positive on average; and (b) the spread of forecasts among analysts is significantly smaller on average than the forecast error, i.e. the typical spread of forecasts around the real earning. In other words, analysts' forecasts are more similar to each other than they are to the variable they are trying to forecast. More recently, a similar data set has been studied to estimate the fraction of herding analysts \cite{Lim}, with the surprising conclusion that around 75\% of the analysts in the data set displayed a marked tendency to herd as a forecasting strategy. Interestingly, about 10\% of analysts were instead found to be anti-herding. Stock prices in turn tend to react more strongly to forecasts that differ from the consensus (see e.g. \cite{Jega} for a recent study).

While psychological factors like social pressure, reputation issues and (for anti-herding) a desire for visibility can be crucial in determining this scenario, it is quite difficult to explain these results within the efficient market hypothesis (EMH). In the EMH world, one would expect different forecasters to use their respective partial information to produce a proxy for the target variable that is unbiased and such that the forecast dispersion and the forecast error are roughly the same, as would result from forecasters that are independent and fully heterogeneous with respect to information and forecasting ability. 

Different models in the economic literature have addressed the problem of the origin of herding behavior among financial forecasters in a Bayesian game theoretic setting (see e.g. \cite{Lim,trueman}), proving that if an analyst aims at maximizing the value of his reputation with investors (or the chances that investors believe he's a good forecaster) then it may actually be more profitable for him to replicate other agents' forecasts rather than putting forward a guess based on his private information. A recent agent-based model proposed by Curty and Marsili \cite{curty} focuses instead on the limits that herding imposes on the efficiency with which information is aggregated. Specifically, it was shown that when the fraction of herders in a population of agents increases, the probability that herding produces the correct forecast (i.e. that individual information bits are correctly aggregated) undergoes a transition to a state in which either all herders forecast rightly, or no herder does.

In this note we study a variation on the theme by Curty and Marsili, aiming at characterizing further the dynamical interplay between learning and heterogeneity of information in a population of agents aiming to predict a continuous exogenous random variable (learning was briefly considered in a discrete forecasting setting also in \cite{curty}). At each time step, every agent is required to formulate a forecast either using his private information or by herding with a group of peers and selects the strategy to adopt based on his past performance. We show that the structure of the agents' choices changes significantly depending on the heterogeneity of information. In particular, herding becomes increasingly preferred by agents as information becomes more and more unevenly spread across the population. However, the herding coefficient (measured by the ratio of the forecast error to the forecast dispersion) peaks roughly where informational inhomogeneity is maximal, implying that learning in such conditions does not allow for an efficient aggregation of the available information.

The results we discuss are mostly obtained by computer simulations. Deeper analytical progress (beyond the simple considerations made here) could be possible either along the lines of \cite{veglio} or by reasonably simplifying the coupled herding and learning mechanisms.

\section{Model definitions}

Following \cite{curty,deffuant}, we consider a population of agents (labeled $i=1,\ldots,N$) who have to forecast  at each time step $t=1,2,\dots,$ a continuous random variable $E$ drawn independently at each $t$ from a uniform distribution in $[0,1]$. The forecast $f_i(t)$ of agent $i$ at time $t$ is taken to be correct if $|f_i(t)-E(t)|<\epsilon$, in which case we shall write $f_i(t)=_{\epsilon} E(t)$. In what follows, the resolution parameter $\epsilon$  will be fixed to the value $0.1$ so as to focus the study on the remaining parameters. In order to fomulate his forecast, every agent must choose between using his private information (strategy lebeled $p$) and herding (label $h$). In the former case, agents simply propose a forecast $f_i^p$ which is correct with probability  $p_i\geq 2\epsilon$, initially unknown to $i$. Note that $2\epsilon$ is the probability with which a random uniform guess in $[0,1]$ is within $\epsilon$ from $E$, hence we are assuming that the private information has better-than-random predictive power. Larger values of $p_i$ correspond to higher forecasting abilities. We assume, along the lines of \cite{curty}, that the  $p_i$'s are sampled independently from
\begin{equation}
\ell(p)=\beta (1-2\epsilon)^{-\beta}(1-p)^{\beta-1} \quad,\quad p\in[2\epsilon,1]\label{distr}
\end{equation}
Tuning $\beta$ one passes from a situation in which almost all agents are well informed (small $\beta$), to one in which almost all agents have no forecasting ability (large $\beta$); as $\beta$ increases, the information heterogeneity (or the a priori forecasting ability) reaches a maximum when $\beta=1$, corresponding to a uniform distribution. We shall denote by $\overline{p}$ the average value of $p_i$, given by $\overline{p}=(1+2\epsilon\beta)/(\beta+1)$.

When herding, an agent uses instead a prediction $f_i^h(t)$ obtained by pooling a group $\mathcal{P}_i$ of $K$ peers chosen randomly and uniformly for each $i$ (our results do not appear to depend significantly on $K$ as long as $K$ is not extensive; we shall use $K=10$ here). This represents, in analogy with e.g. \cite{angh}, the ``contact network'' of the agent. Our choice for a plain Erd\"os-Renyi topology parallels that made in \cite{curty}, but results are expected to change if different topologies are employed. The herding forecast is defined as the fixed point of the iterative process
\begin{equation}
f_i^h(n+1) = \frac{f_i^h(n) + \sum_{j\in \mathcal{G}_i(t)} f_j(n)}{|\mathcal{G}_i(t)| + 1}\label{herd}
\end{equation}
with $\mathcal{G}_i(t) = \{j \in \mathcal{P}_i : f_i^p(t) =_d f_j^p(t) \}$. In words, at time step $t$ every agent only interacts with his peers whose initial (private) guess is sufficiently close, within a range measured by $d$, to his \cite{deffuant}. The ``effective interaction range''  is a crucial parameter to model social interaction and has proved to play a non trivial role in other contexts \cite{cas,col}. Note that the number $|\mathcal{G}_i(t)|$ of such peers is obviously bounded by $K$ but it fluctuates in time; furthermore, in the above sum $f_j(n)$ denotes the forecast of agent $j$ who may be herding (in which case $f_j$ changes with $n$) or not (in which case it is fixed to $f_j^p$). This defines a dynamical ``imitation network'' onto the contact network, close in spirit to that defined in \cite{angh} in the context of Minority Games. In this case, however, the averaging operation through which herding is performed does not allow for a straightforward identification of an imitation hierarchy.

As in many other instances of games with learning \cite{cmz}, we take agents to be inductive: they monitor the performances of their two strategies over time via scoring functions indexed by $g\in\{p,h\}$ that are updated in time according to
\begin{equation}
U_i^g(t+1) - (1 - \lambda)U_i^g(t) = \pi_i^g(t) - \delta_g
\end{equation}
and the agent's chosen strategy $g_i(t)$ is selected by a logit rule with learning rate $\Gamma\geq 0$:
\begin{equation}
\text{Prob}\{g_i(t)=g\}= \frac{e^{\Gamma U_i^g(t)}}{e^{\Gamma U_i^p(t)}+e^{\Gamma U_i^h(t)}}~~,~~~g\in\{p,h\}
\end{equation}
The different parameters appearing above have all been introduced and discussed at length in the context of Minority Games and related models (see e.g. \cite{rev} for a broad review). In the present case, $0\leq\lambda\leq 1$ denotes a memory length parameter, roughly corresponding to the inverse of the time scale over which agents preserve a memory of the past performance of their strategies. $\delta_{p,h}$ denote instead the cost (or the incentive) faced by each player to get his private information or to herd. $\pi_i^{p,h}(t)$ denotes the profit faced by agent $i$ at time step $t$. For the sake of simplicity, we reward agents who guess correctly with one point, while we take one point from agents who guessed incorrectly. Finally, $\Gamma$ is a parameter that encodes for a tunable stochasticity in the agents' choice rules, with deterministic behavior recovered for $\Gamma\to\infty$. Note that at every time step both the strategy scores of every agent are updated.

Starting from initial conditions $U_i^g(0)=0$, we are interested in observing the steady state behavior of the following quantities:
\begin{itemize}
\item The herding probability, measured by the time-averaged fraction of herders:
\begin{equation}
F_h = \avg{\frac{1}{N}\sum_{i}\delta_{g_i(t),h}}_t
\end{equation}
\item The probability of success of strategy $h$ (averaged over time and agents):
\begin{equation}
q=\avg{\frac{\sum_{i}\delta_{g_i(t),h}~\chi(f_i^h(t)=_\epsilon E)}{\sum_{i}\delta_{g_i(t),h}}}_t
\end{equation}
where $\chi(A)=1$ if the event $A$ is true, and zero otherwise
\item The time- and agent-averaged forecast error:
\begin{equation}
\Sigma = \avg{ \frac{1}{N}\sum_{i}(f_i(t) - E(t))^2}_t
\end{equation}
\item The time-averaged forecast dispersion:
\begin{equation}
\sigma = \avg{ \frac{1}{N}\sum_{i}(f_i(t) - \bar{f}(t))^2}_t
\end{equation}
\item The {\it herding ratio } $\phi=\Sigma/\sigma$
\end{itemize}
The way in which agents produce their forecasts, and as a consequence the herding ratio will be affected by all of the above parameters but, most importantly, will depend on the information heterogeneity, measured by $\beta$. Note that $\phi\geq 1$. We remind, finally, that, under the EMH, $\Sigma\simeq\sigma$, so one would expect $\phi\simeq 1$. 

\section{Results}

\subsection{The simplest case}

\begin{figure}
\begin{center}
\includegraphics[width=10cm]{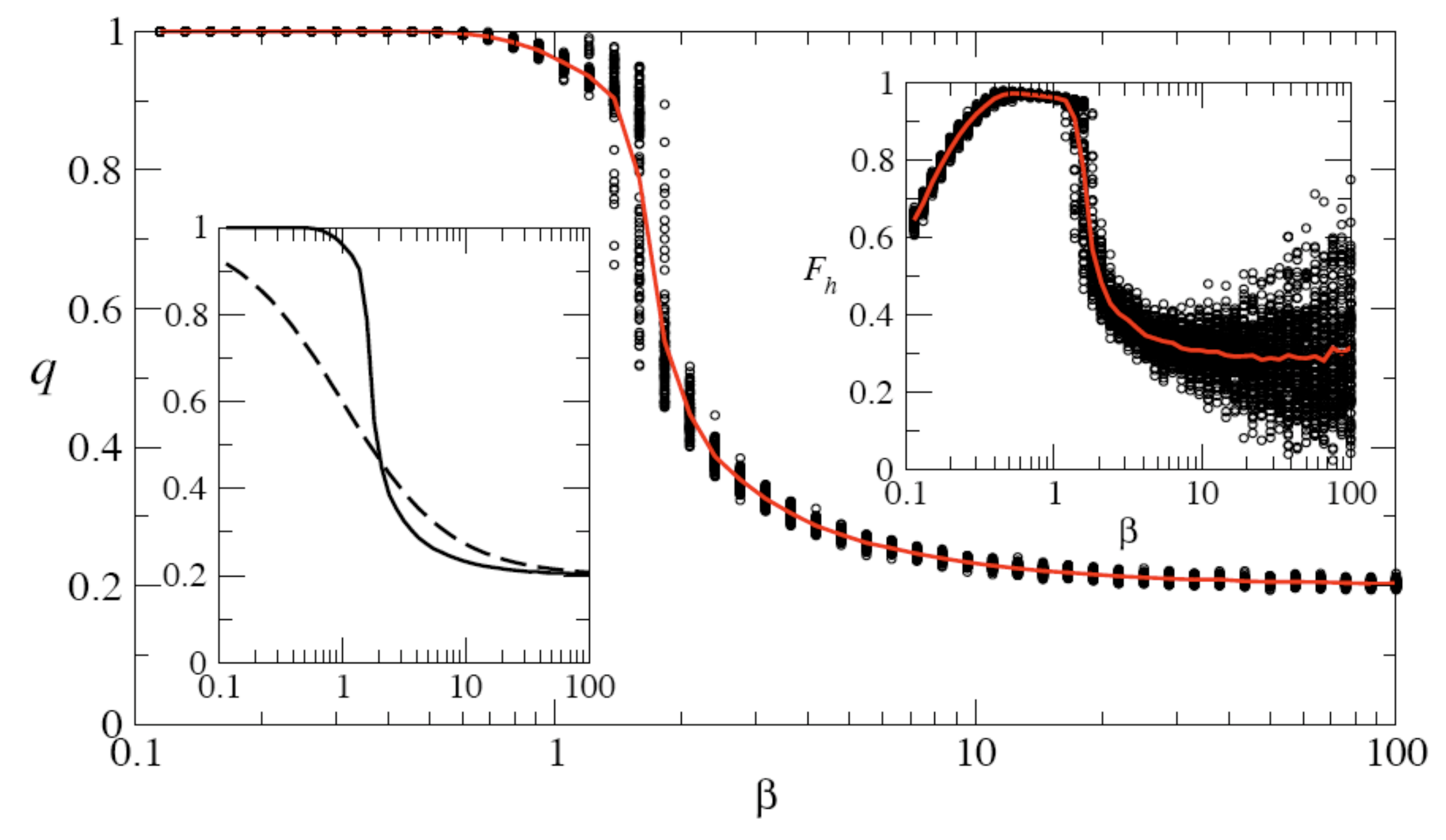} 
\caption{\label{uno}Probability $q$ of successful herding versus $\beta$. Markers denote results for individual samples; the continuous line stands for the average. Left inset: $\overline{p}$ (dashed line) and $q$ (continuous line) versus $\beta$. Right inset: $F_h$ versus $\beta$. Simulation parameters: $N=1000$, $K=10$, $\epsilon=0.1$.} 
\end{center}
\end{figure}

\begin{figure}
\begin{center}
\includegraphics[width=10cm]{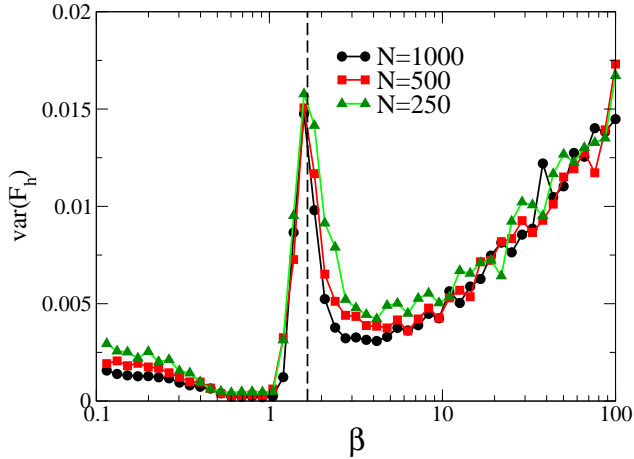} 
\caption{\label{fluc}Sample-to-sample variance of $F_h$ versus $\beta$ for different values of $N$. Simulation parameters: $K=10$, $\epsilon=0.1$. The dashed vertical line marks the position of $\beta^\star=5/3$ in this case. }
\end{center}
\end{figure}

\begin{figure}
\begin{center}
\includegraphics[width=10cm]{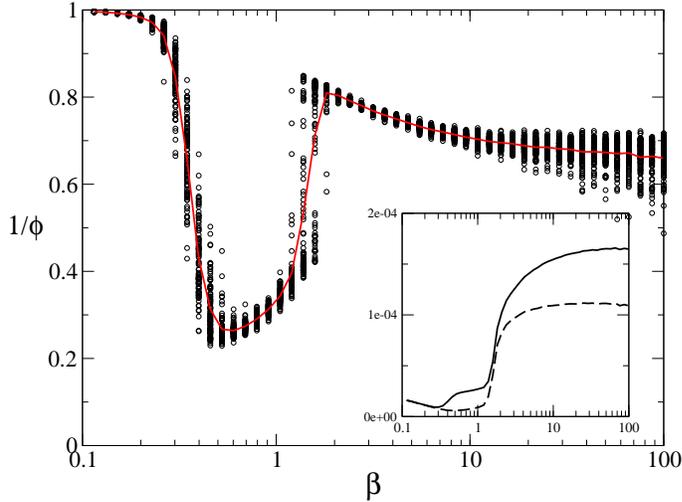} 
\caption{\label{due}Inverse herding ratio $1/\phi$ versus $\beta$. Inset: $\Sigma$ (continuous line) and $\sigma$ (dashed line) versus $\beta$. Simulation parameters: $N = 1000, K = 10$, $\epsilon=0.1$.} 
\end{center}
\end{figure}

We begin by analyzing the case in which every agent consults all his peer group ($d=1$ or $\mathcal{G}_i(t)=\mathcal{P}_i$ for all times), has infinite score memory ($\lambda=0$) and learning rate ($\Gamma=\infty$, corresponding to a deterministic choice rule) and faces no costs (or receives no incentives) to use his strategies ($\delta_h=\delta_p=0$). This will be used as a reference situation to evaluate the impact of the above parameters on the game. Fig.~\ref{uno} shows the average herding success probability $q$ as well as the fraction of herders $F_h$ as a function of $\beta$. 
We see that  when agents are well informed ($\beta$ small) herding outperforms the private information strategy and the majority of agents correctly learn to herd to increase their success probability well above $\overline{p}$. For large $\beta$, on the other hand, when agents have limited predictive ability, they learn to use their private signal as it slightly outperforms the herding forecast. This behavior, including the decay of $q$ that is observed for large $\beta$ ($q\to 2\epsilon$ as $\beta\to\infty$) essentially parallels that observed in \cite{curty}. The behavior of $F_h$, by contrast,  displays a sharp drop as informational heterogeneity increases beyond the point where, as is clear from the inset, the private signal outperforms herding. Note that for large $\beta$ sample-to-sample fluctuations become larger and larger as $\overline{p}\simeq q$, so that realizations with many herders (up to a 70\% fraction of the population) coexist with realizations with few herders (less than 10\%). It is worthwhile to inspect the dependence of sample-to-sample fluctuations on $\beta$ more closely, see Fig. \ref{fluc}. 
Besides the increase for large $\beta$, a sharp peak (roughly independent of the system size) appears for intermediate $\beta$, marking a qualitative change in the game's macroscopic properties. Again, this signals a strong sample-dependence of the fraction of herders and occurs when the payoffs of the two strategies become comparable. Such an effect is absent in the one-shot game and is induced by learning. Note that the fraction of herders for $\beta$'s  close the the peak is consistent with the results of \cite{Lim}. In turn, the herding ratio displays the behavior shown in Fig. \ref{due}.
One observes a sharp minimum in $1/\phi$ around $\beta\sim1$, where it attains a value consistent with that found empirically in \cite{bouchaud,Lim}, where $\phi\sim 5$. Away from this crossover region (where, we remind, information heterogeneity is maximal), $\phi$ is instead closer to the EMH limit $\phi\sim 1$. This suggests that inductive agents indeed manage to aggregate information quite efficiently when it is distributed uniformly across them, regardless of the quality of the information. This process is no longer possible in presence of a more diversified information landscape.

A naive analytical estimate of the value of $\beta$ where the game undergoes a crossover can be obtained by arguing as in \cite{curty}. Denoting by $\pi$ the probability of a correct forecast, one has
\begin{equation}\label{a}
\pi = F_h q + (1-F_h)\bar{p} 
\end{equation}
where $q$ denotes the probability of a correct forecast by herding. Neglecting both the learning and the herding dynamics, $q$ is given by the probability that the majority of the peer group members have a correct forecast, i.e.
\begin{equation}\label{b}
q =\sum_{g = K/2+1}^{K+1} \binom{K+1}{g}\pi^g(1-\pi)^{K+1-g}
\end{equation}
A qualitative change is expected to occur when $p=\overline{q}=\pi$. One easily sees from (\ref{b}) that this condition is satisfied (besides the trivial solutions $\pi=0$ or $\pi=1$) when $\pi=1/2$, implying
\begin{equation}\label{bstar}
\beta^\star=\frac{1}{1-4\epsilon}
\end{equation}
For $\epsilon=0.1$ the prediction $\beta^\star=5/3$ agrees well with the numerical experiments. Eq. (\ref{bstar}) also characterizes the role of the resolution parameter: as $\epsilon$ increases, the number of effective alternative ``states'' of the variable $E$ (given by $1/(2\epsilon)$) decreases and $\beta^\star$ becomes larger, extending the range of effectiveness of herding as a forecasting strategy. (The fact that $\epsilon$ cannot exceed $1/4$ was implicit in the model's definition.)

Note that the na\"ive  guess for the fraction of herders given by  \cite{curty}
\begin{equation}
F_h = \int_{2\epsilon}^{\bar{p}}\ell(p)dp=1-\left(\frac{\beta}{1+\beta}\right)^\beta
\end{equation}
according to which agents with $p_i<\overline{p}$ are assumed to herd asymptotically, does not produce the correct results in the present model, suggesting that the specific forms of $F_h$ and of its fluctuations, including the large $\beta$ increase, are likely due to the learning dynamics itself.

\subsection{Role of the parameters $\Gamma$, $\lambda$, $\delta_h$, $\delta_p$ and $d$}

\begin{figure}
\begin{center}
\includegraphics[width=12cm]{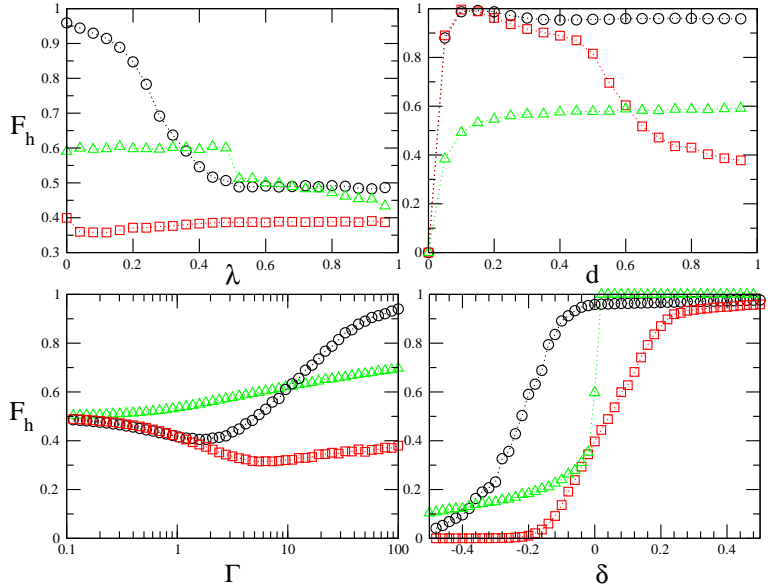} 
\caption{\label{param}Fraction of herders $F_h$ versus $(\lambda, d, \Gamma, \delta=\delta_p-\delta_h)$ for $\beta=0.1$ (triangles), $\beta=1$ (circles) and $\beta=3$ (squares). Simulation parameters: $N = 500, K = 10$, $\epsilon=0.1$.} 
\end{center}
\end{figure}

\begin{figure}
\begin{center}
\includegraphics[width=12cm]{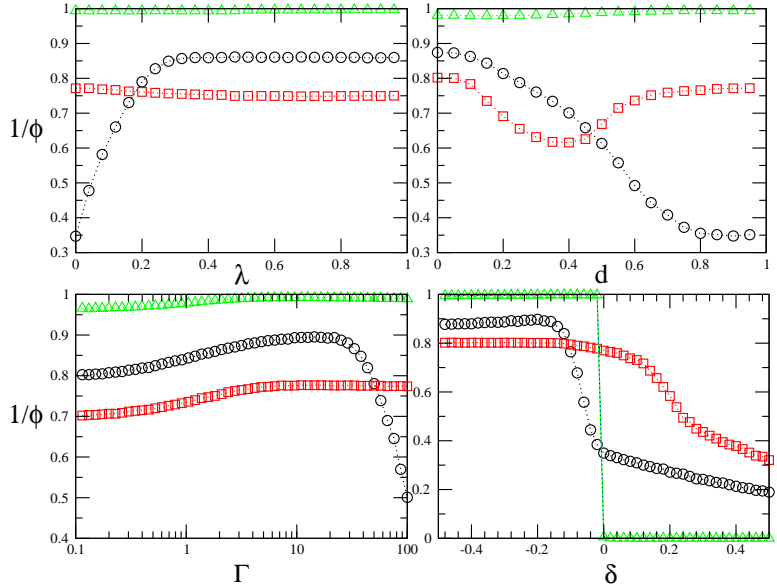} 
\caption{\label{param2}Inverse herding ratio $1/\phi$ versus $(\lambda, d, \Gamma, \delta=\delta_p-\delta_h)$ for $\beta=0.1$ (triangles), $\beta=1$ (circles) and $\beta=3$ (squares). Simulation parameters: $N = 500, K = 10$, $\epsilon=0.1$.} 
\end{center}
\end{figure}

We now focus our analysis on the additional model parameters described in Sec. 2. For the sake of simplicity, we will study one of them at a time as variations of the basic case investigated above. Figures \ref{param} and \ref{param2} show specifically how the fraction of herders and the herding ratio change when these parameters are varied for different values of $\beta$. 

Starting with $\lambda \in [0,1]$, which as said above introduces a finite memory of past scores (for $\lambda=0$ there is no loss of memory while for $\lambda>0$ the score is exponentially discounted in time with a characteristic time given by $1/\lambda$, see \cite{cdm} for a solvable model that highlights its role in the context of Minority-like Games), we see both $F_h$ and $\phi$ are essentially insensitive to $\lambda$ except in the ``critical region'' where informational heterogeneity is maximal. In this case, a larger $\lambda$ (or a shorter memory) leads to a decrease of both $F_h$ and $\phi$, implying that a finite memory may lead closer to the rational aggregation of the available information when the latter is distributed in a highly in-homogeneous way.

Similar conclusions can be reached by analyzing the learning rate $\Gamma$ that tunes the amount of stochasticity in agents' choice. Its impact appears to be most remarkable close to $\beta=1$, when a (small) degree of randomness in the strategy selection process helps to reduce the fraction of herders and hence to bring $\phi$ closer to the rational value of $1$. 

The role of parameter $d \in [0,1]$ in the context of opinion dynamics models has been discussed e.g. in \cite{deffuant}. It takes into account the chance that people may prefer to interact, and possibly accept to adjust their beliefs, only with opinions not too far from their initial ones. For $d=1$ all opinions in the peer group are treated equally. For other values of the parameter, instead, the sum in \eqref{herd} is restricted only to agents whose forecasts are within $d$ from each agent's initial choice. It is interesting to note that when herding represents the best strategy, i.e. around $\beta=1$, the fraction of herders is almost insensitive to different values of $d$ (which possibly suggests also a weak dependence of the results on $K$). Nevertheless reducing $d$ the herding ratio appears to get smaller, indicating that  a more rational aggregation of information may occur. By contrast, for larger $\beta$ agents may be unable to identify $p$ as the most likely successful strategy if $d$ is too small.

Coming finally to the incentives $\delta_h,\delta_p$ (which are known to have a far from trivial impact on Minority Games, see e.g. \cite{cm}), we focus on the dependence of $F_h$ on the parameter $\delta=\delta_p-\delta_h$, which is easily understood to be the relevant quantity in this case. As to be expected, incentives to herding, or higher costs for using the private signal, shift agents towards the $h$ strategy, and lead to an increase of $\phi$ (and vice versa for incentives to the $p$ strategy). This crossover appears to be smooth only when $\beta$ is sufficiently large. For smaller $\beta$ it sharpens and one observes a steep jump at $\delta=0$ in both $F_h$ and $\phi$, reminiscent of similar effects induced by incentives or Tobin taxes in Minority Games \cite{cm,gal}. In this case, agents appear to polarize on the herding strategy as soon as a small incentive is available, leading to disastrous consequences for $\phi$ and suggesting the existence of a first-order transition in $\delta$ (though a more refined numerical analysis would be needed to clarify this point). The qualitative outlook is however essentially unchanged with respect to the the case of larger $\beta$'s.

\section{Conclusions}

We have studied here a simple forecasting game with inductive agents who must formulate a forecast by either using their private information or by herding with a group of peers. The quality of the information at an agent's disposal is measured by the a priori predictive ability of his private signal, and we investigate how the game's overall properties are affected by increasing informational heterogeneity in the population. Our main result is that inductive agents may be unable to produce rational forecasts when the heterogeneity is large. In this situation, the herding ratio becomes significantly larger than one, taking on values similar to those measured empirically in the financial literature \cite{bouchaud,Lim}. We have also observed that the efficacy of herding depends strongly on the distribution of information. The role of several parameters of interest in the context of games with inductive agents has finally bee analyzed. Generically speaking, a finite learning rate, a shorter memory or a smaller interaction range may all contribute to reduce the herding ratio when the informational heterogeneity is large.

The interest in forecasting games is based on the fact that they present a simple outlook and a rich phenomenology that allows to shed some light on the process of information aggregation and its limits. It is however difficult to establish a direct contact between these toy models and financial markets, which still represent the main source of empirical data. One possible step in this direction that is worth exploring would consist in coupling the event to be forecasted, $E$ here, with the agents' choices, as done for example in \cite{JDF}. $E$ would play in these models the role of a `price'. Depending on the form of the payoff one would then observe a dynamical feedback between learning and `price' leading to a rich outlook possibly similar to that studied in Minority Games.





\end{document}